\documentclass[conference]{IEEEtran}
\IEEEoverridecommandlockouts
\usepackage{cite}
\usepackage{amsmath,amssymb,amsfonts}
\usepackage{algorithmic}
\usepackage{graphicx}
\usepackage{textcomp}
\usepackage{xcolor}
\usepackage{hyperref}
\def\BibTeX{{\rm B\kern-.05em{\sc i\kern-.025em b}\kern-.08em
    T\kern-.1667em\lower.7ex\hbox{E}\kern-.125emX}}
\begin{document}

\title{A Goal-Oriented Chatbot for Engaging the Elderly Through Family Photo Conversations}

\author{
\IEEEauthorblockN{Raymond Chung, Keith Ng, CD Shum}
\IEEEauthorblockA{
\textit{Logistics and Supply Chain MultiTech R\&D Centre} \\
Hong Kong, China \\
rchung@lscm.hk, kng@lscm.hk, cdshum@lscm.hk
}
}

\maketitle
\begin{center}
    \vspace{-1em}
    \small
    © 2025 IEEE. This is the author’s version of the work. 
    The final published paper is available at 
    \href{https://doi.org/10.1109/COMPSAC65507.2025.00282}{DOI: 10.1109/COMPSAC65507.2025.00282}.
\end{center}
\begin{abstract}


We propose a personalized chatbot designed for elderly individuals. The chatbot initiates discussions based on family photos, encouraging users to interact naturally.  During these interactions, it generates W questions—who, where, when, and what—to stimulate cognitive function, followed by an open-ended question to promote positive reminiscence. This approach is structured around a goal-oriented dialogue framework. Additionally, after each conversation about a photo, the chatbot analyzes the discussion to identify topics that the user favors or dislikes. It then offers the user the option to chat about another photo either featuring the same family members or an individual previously mentioned in the conversation. To support this system, we have developed a web portal that allows caregivers to upload photos and review chat conversations. This personalized chatbot not only encourages elderly users to engage with the chatbot regularly and reduces feelings of loneliness but also provides caregivers with a valuable tool to gain insights into users' well-being. 


\end{abstract}


\section{Introduction}

As individuals age, they often experience a decline in social connections, which can lead to feelings of loneliness. The COVID-19 pandemic further intensified social isolation among older adults due to prolonged lockdowns and restricted interactions. Engaging in positive reminiscence can foster a resilience mindset, helping them better cope with the psychological stress of social isolation\cite{b0}. While caregivers play a crucial role in alleviating loneliness by discussing family-related topics and encouraging positive reflection, their availability is limited. As the elderly population continues to grow, the demand for caregiving support is increasing, highlighting the need for scalable and sustainable solutions.

Despite the widespread use of social media platforms and instant messaging among younger generations, elderly individuals may still face obstacles that hinder their social interactions. Poor digital literacy skills can make it challenging for them to use applications like WhatsApp or Facebook. Additionally, vision problems and reduced dexterity can complicate reading small text and navigating buttons on screens. In this context, chatbots, which are computer programs designed to mimic human conversation, can offer speech-based interaction. Hence, for elderly individuals who lack of physical social interactions, chatbots can serve as a crucial source of companionship and engagement.


Not all chatbots are well-suited for elderly users. To ensure meaningful engagement, a personalized chatbot tailored to their interests is essential. For example, a chatbot that encourages users to reminisce about recent events and daily routines can foster deeper interaction. Additionally, initiating discussions about family members can further motivate elderly users to participate in conversations.

Recently,  large language model (LLM) technology, e.g. GPT\cite{b01}, has become popular for enhancing the chatbot’s ability to provide natural and intuitive interactions. As an advanced language model, GPT generates human-like responses, allowing the chatbot to adapt dynamically to users’ preferences and create engaging, personalized conversations.



The proposed chatbot system uses GPT technology to facilitate meaningful conversations about personal photos and past experiences. It includes a web portal for caregivers, and a mobile app for elderly users to interact via speech. The chatbot features a dialogue agent to manage chat flow. It enhances user engagement, encouraging elderly individuals to reminisce, socialize, and maintain cognitive stimulation.



\section{Related Work}

Dialogue systems, or chatbots, can be broadly classified into two categories: goal-based and open-domain.

Goal-based dialogue systems are designed to perform specific tasks efficiently. For example, Liu\cite{b65} developed a machine model for detecting intent from speech and tested its effectiveness in client air flight booking. Similarly, Cao\cite{b7} proposed a chatbot tailored for medical consultations, demonstrating its potential in healthcare applications.

Open-domain dialogue systems, on the other hand, focus on engaging users in free-flowing, natural conversations without predefined objectives. These chatbots serve as companions, providing entertainment and facilitating casual discussions across a wide range of topics \cite{b8}.

While goal-based chatbots excel in structured, task-oriented interactions, open-domain chatbots enhance engagement and personalization. 


GPT technology has emerged as a cutting-edge solution for developing dialogue agents. Ham Donghoon et al. \cite{b5} proposed a structured approach for building a goal-based dialogue system utilizing GPT technology. Their method involves designing a dialogue state to record essential information related to a goal and implementing system actions based on both the dialogue state and conversation history. By training a GPT-2 model, they significantly improved the success rate of achieving goal-oriented outcomes, thereby enhancing the overall performance of the dialogue system.

Alessa et al\cite{b1} proposed the development of a chatbot specifically tailored for the elderly. Initially, the caregiver is required to complete a detailed questionnaire, which includes demographic information, personality traits, hobbies, and daily activities of the elderly person. Instead of training a model, all this information is described in the system prompt of GPT3 to ensure a personalized interaction experience. Furthermore, the proposed chatbot features a quiz mode, in which it engages the elderly with general knowledge questions, and a health mode, allowing the elderly to report any health concerns to the chatbot. However, the effectiveness of this approach is limited by the static nature of the background information provided by caregivers, which does not capture dynamic, up-to-date details about the elderly individual's recent experiences or evolving interests. In contrast, our chatbot continuously adapts conversations, incorporating new photos and descriptions provided by family members.  

Tao et al\cite{b2} proposed a proactive chatbot that focuses on visual question generation. They developed a dataset where each image includes a question topic and a question object. For example, if an image shows a dog jumping off a dock, and the topic is hobbies and the object is the dog, the generated question might be, "Do you have a pet? If so, do you take it to outdoor activities like jumping off a dock?" Alternatively, if the topic is about travel experiences, the question would ask whether the elderly person has met a smart and funny dog during their travels.
Their main focus is to train a model to correctly extract topics from images and then measure the similarity between the generated questions and the pre-written ones. Our chatbot also focuses on photo-based conversations, specifically on family-related photographs. This approach serves to create a more engaging conversation with the user.

Collecting an appropriate dataset for a companion chatbot development presents significant challenges. Chen et al\cite{b3} proposed a chatbot designed to emphasize empathy, active listening, and providing comfort. The research involved the collection of a substantial corpus of single-turn, long-text psychological counseling conversations, encompassing over 200,000 counseling questions and 600,000 answers obtained via data outsourcing services. Subsequently, a LLM was employed to transform these interactions into conversations that meet the empathy requirement. After filtering the data, the researchers curated a dataset of over 2 million samples, which was then used to fine-tune the chatbot. 

Jiang et al\cite{b4} developed a conversation dataset grounded in principles from cognitive simulation handbooks. This dataset comprises 2,600 samples, with each sentence annotated with three distinct labels: a cognition label (e.g., respect, conformity, reminiscence), an emotion label (e.g., happiness, liking, sadness), and a strategy label (e.g., questioning, self-disclosure, providing suggestions). An LLM model is fine-tuned with this small set of data. These labels are predicted based on the conversation content, and the generated responses are conditioned on this predicted label information. 

Instead of fine-tuning an LLM with extensive data, we use a dialogue agent to manage conversation flow while relying on an LLM for natural response generation. To refine interactions, we simulate conversations between an elderly persona LLM and our chatbot, carefully crafting prompts to ensure responses are friendly, engaging, and supportive. This prompt-engineered approach enables dynamic, natural interactions without requiring large-scale model training, making it an efficient and adaptable solution for elderly companionship.

\section{Methodology}

Our proposed chatbot features a dialogue agent designed to foster meaningful interactions with elderly users. It leverages the user's profile, photo records, and conversation history to generate engaging and contextually appropriate responses, ensuring a personalized and dynamic conversational experience.

\subsection{Personalizing the Chat}

The system leverages the elderly user's background and profile. The background includes basic personal details, while the profile outlines the user's likes and dislikes. Caregivers create an account for the user, providing basic background details and uploading photos of family members. When new images are added, the system automatically detects and recognizes familiar faces. Caregivers can then input descriptions for each photo. When engaging with the user, the chatbot leverages the profile and family photos to generate personalized, contextually relevant discussions, ensuring interactions are closely related to the user.

\subsection{Engaging the User's Interest}

The chatbot initiates conversations with a family photo. The system generates a series of engaging questions based on the provided descriptions, following a goal-based dialogue approach to maintain a structured conversation while emphasizing key details of the event captured in the image. This method encourages users to share and reminisce about meaningful past experiences. Throughout the chat, if the user elaborates on a specific aspect of the photo, the chatbot engages with the new topic before guiding the conversation back to the original subject, ensuring a natural and user-centered interaction. After discussing a photo, the chatbot prompts the user to explore additional photos featuring the same family member or the member the user just mentioned, fostering a more coherent and emotionally meaningful conversation centered around loved ones rather than isolated photo events.

\subsection{Cognitive Stimulation}

To support cognitive engagement, the chatbot gently provides corrective feedback when a user gives an incorrect answer to a question. This approach encourages users to actively recall information and refine their memory through interactive dialogue. Additionally, our system functions as a dynamic, evolving album designed to facilitate ongoing interactions. When new photos are uploaded by caregivers, the chatbot prioritizes conversations around these fresh images, ensuring discussions remain relevant and engaging. In the absence of new photos, the chatbot selects older images that have not been discussed for a prolonged period, prompting the user to revisit past memories. By strategically cycling through both recent and older photos, the system fosters cognitive stimulation, helping users maintain memory recall and connection to their personal experiences.

\subsection{Caregiver Summary}

At the conclusion of each interaction, the system updates the elderly individual's profile based on the conversation history, ensuring a continuously personalized and adaptive engagement experience. Caregivers can access a dedicated web portal to review the conversations linked to each photo, along with a concise summary of key discussion points. This feature enables caregivers to monitor engagement patterns, identify topics of interest, and provide insights on users' well-being. 

\section{EXPERIMENTS AND RESULTS}

\subsection{Web interface}
We developed a web interface to enable caregivers to input essential information about the elderly and upload relevant photos. Caregivers provide images of family members, which are then used for face verification. Upon uploading a new photo, our system employs face recognition technology to detect whether any familiar family members are present in the image. For this step, we utilize the pre-trained face recognition library\footnote{https://github.com/ageitgey/face\_recognition}. Once the system verifies the faces, caregivers can add descriptions for each photo to enrich the chatbot's conversational context. Additionally, users can upload text-based conversations from a messaging app, serving as an alternative source of discussion when no new photos are available. These messages provide valuable topics for initiating conversations with the elderly, ensuring engagement even in the absence of visual content. The structure and capabilities of the web interface are visually summarized in Fig. 1.

\begin{figure}[tbp]
    \centering
    \includegraphics[width=0.45\textwidth]{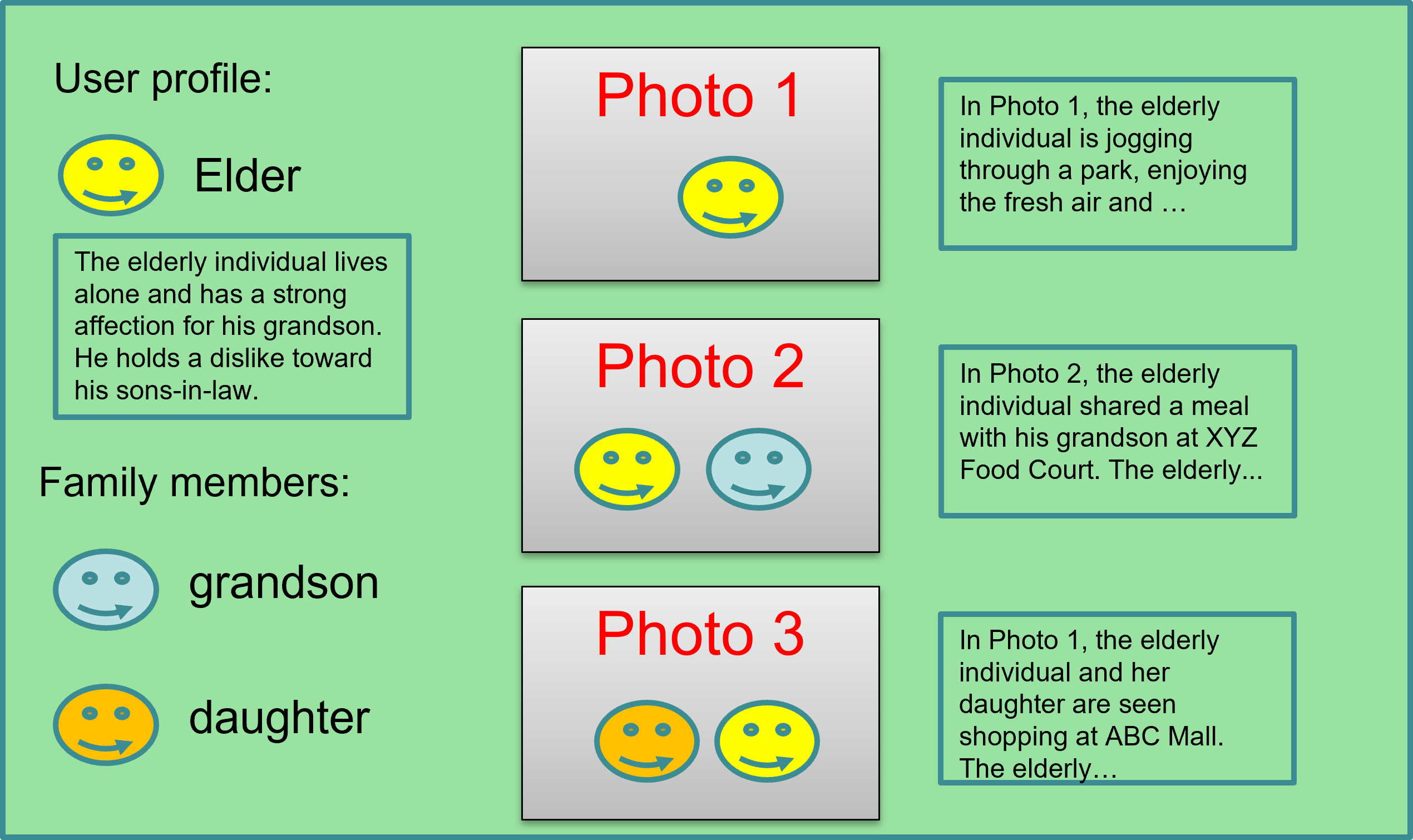}
    \caption{Web portal for adding a new user and the related photos}
    \label{fig:ourtts}
\end{figure}

\subsection{Question and Answer Generation}
To facilitate meaningful conversations, our system employs GPT to generate personalized questions for each uploaded photo. These questions are designed to encourage elderly users to recall past events, fostering memory engagement and reminiscence. The chatbot avoids abstract or overly complex topics, instead focusing on concrete details such as location, time, activities, and individuals present in the image. The prompt used for question and answer generation is shown in Table I.

\begin{table}[h]
\centering
\begin{tabular}{|l|}
\hline
\textbf{GPT Prompt} \\ 
\hline
As a social worker, you need to have a friendly conversation with an elder.\\
I will give you a description of an existing photo, and you will provide\\
 questions and answers. \\
The description of the photo is as follows: \{image\_description\}. \\
Generate simple questions that encourage the elderly to recall details \\
on their own. Your questions should follow this sequence when \\
discussing the photo: \\
1. Where was it taken?\\
2. When did the event occur?\\
3. What is happening in the image?\\
Use the provided photo description to create relevant question-answer\\ pairs. Format responses as follows: Question\#\#\#Answer;\\
\hline
\end{tabular}
\caption{Prompt for Question and Answer Generation}
\label{tab:qa_prompt}
\end{table}

The questions and answers are stored as a list object within the chatbot program, allowing it to reference them during conversations with the user. Furthermore, we incorporate two additional questions. The first question identifies who is in the photo, with the answer derived from face recognition results. The final question invites the elderly user to share any thoughts or memories related to the photo, fostering open-ended conversation and personal reflection.

Once questions are generated, the chatbot engages the user in conversation, presenting each question in a friendly and interactive manner. As the user responds, the system evaluates their answers and provides gentle corrections when necessary, reinforcing learning and memory retention in a supportive way.



\subsection{Dialogue Agent Flow Control}

Following the approach outlined in \cite{b1}, we define the role of GPT-4 using the following prompt shown in Table II.

\begin{table}[h]
\centering
\begin{tabular}{|l|}
\hline
\textbf{GPT Prompt} \\ 
\hline
You are a social worker and you need to have friendly and\\
encouraging conversations with an elder. The whole conversation\\
should be conducted in a relaxed atmosphere and you should \\
not repeat what the elder has said. \\
The elder's background is as follows: \{background\}, \{profile\}.\\
You and the user are viewing a photo together. \\
The description of the photo is as follows: \{image\_description\}. \\
\hline
\end{tabular}
\caption{GPT role description for conversational interactions}
\label{tab:role_description}
\end{table}

Then, we instruct the chatbot to ask a question. The first question will be asking whom is on the photo. Then the elder will give a reply to the chatbot. 

Rather than functioning solely as a conversational companion, our chatbot is designed as a Dialogue Agent, meaning it must adhere to specific rules to effectively manage the flow of conversations. During interactions, the chatbot continuously monitors user input and selects the appropriate response strategy:

\begin{itemize}
    \item If the user answers a question correctly, the chatbot acknowledges their response positively before proceeding with the next question.
    \item If the answer is incorrect, the chatbot gently corrects them before proceeding with the next question.
    \item If the user shifts away from the expected response, the chatbot engages with their input organically, prioritizing fluid conversation rather than rigid adherence to predefined questions.
    \item If the discussion stagnates or becomes repetitive, the chatbot asks again the unanswered question or transitions to a new photo when no questions left.   
     \item If the user loses interest in continuing the chat, the chatbot politely suggests chatting again next time.

\end{itemize}



The prompt used to achieve this chat flow control is shown in Table III.

\begin{table}[h]
\centering
\begin{tabular}{|l|}
\hline
\textbf{GPT Prompt} \\ 
\hline

The previous question was: \{q\}. \\ 
The correct answer was: \{a\}. \\ 
The user's response is: \{reply\}. \\ 
The next question is: \{q\_next\}. \\

You compare the correct answer and the user's response to see \\ 
if they refer to the same person or time or place.\\

Choose one of the following actions:\\
A: If the user answered correctly, 
respond warmly\\ and ask the next question.\\
B: If their answer was incorrect, gently 
provide the correct answer\\ and proceed 
with the next question, ensuring accuracy.\\
C: If their response does not address 
the previous question,\\ engage with what 
they said naturally instead of rigidly \\
repeating the question.\\
D: If the conversation becomes repetitive, ask again the current\\
question or, if none remain, transition to a new photo. \\
E: If the user shows disinterest in continuing, encourage them \\
to chat again next time. \\
Response format: option:A/B/C/D/E, response: xxxx\\
\hline
\end{tabular}
\caption{Prompt for defining the Dialogue Agent}
\label{tab:flow_control}
\end{table}

We log the sequence of options selected by the dialogue agent and apply constraints to regulate its behavior. During interactions, the chatbot can choose A, B, C, or E. If it selects C, subsequent choices are limited to C, D, or E. When C is chosen too frequently, the available options are further restricted to D or E. When D is selected, the full range of options—A, B, C, and E—becomes available once again. Based on this design, the chatbot follows a structured approach:
\begin{itemize}
\item If the chatbot asks a question and the user responds, it selects A or B to determine whether the answer is correct, then moves on to the next question.

\item If the user does not respond and introduces a new topic, the chatbot selects C to engage in a friendly follow-up.

\item When C is selected too often, the chatbot must then choose D to bring the conversation back to the original question.

\item E remains available at all times as an option to end the chat by the user.
\end{itemize}
This structured framework ensures balanced interactions, keeping the conversation engaging while maintaining conversational flow and relevance.


\subsection{Chat Summary}
Once all pre-defined questions have been asked, the chatbot transitions into open-ended conversation, allowing for more natural and engaging dialogue. When the chat reaches a certain number of exchanges, the chat summary process is triggered.

To optimize personalized interactions, our system updates the user's profile after each chat. It identifies the user's preferred and avoided discussion topics and subjects. Based on this information, the chatbot adapts future interactions to ensure continued engagement and further personalization.

During this chat summary process, the chatbot also detects whether the user has shown a preference for a particular family member. If so, it will offer the option to continue chatting about a photo featuring that individual, fostering deeper connections and coherent discussions.

Additionally, this automated process reduces the need for caregivers to manually update the chatbot’s information. It operates much like a social worker who records new details in a logbook and reviews past entries before visiting elderly individuals, ensuring each interaction is informed, personalized, and supportive. The prompt used to achieve this summary process is shown in Table IV.

\begin{table}[h]
\centering
\begin{tabular}{|l|}
\hline
\textbf{GPT Prompt} \\ 
\hline
Generate a summary of the chat. Based on the user background, previous\\
profile, and chat history, create a new profile that includes:\\
Activities the user enjoys. People the user prefers or avoids.\\
Format: {Like= [], Dislike= []} (each item should be 1-2 words).\\

Inputs:\\
User background: \{background\}\\
User profile: \{profile\}\\
Chat History: \{chat\_history\}\\
Additionally, based on the provided information, suggest who should \\
be featured in the next photo, prioritizing the family member the user \\
discusses most positively.\\
Choose from: \{family members\} \\
Response format: \\
New summary: [Concise chat summary] \\
New profile: YYYY \\
Target Person: ABC \\
\hline
\end{tabular}
\caption{Prompt for chat summary process}
\label{tab:gen_summary}
\end{table}


\subsection{Mobile App Interface}

We developed an Android mobile application and deployed it on a tablet device. The app utilizes Azure Speech-to-Text\footnote{https://azure.microsoft.com/en-us/products/ai-services/ai-speech} to transcribe the user's speech input. The chatbot then generates a corresponding text response, which is subsequently converted into speech using the device's built-in text-to-speech engine.

The user interface of the application is illustrated in Figure 2. Upon clicking the "Start Chatting" button, the app displays a relevant photo. Additionally, a virtual avatar—an animated character appears and synchronizes with the playback of the generated speech, creating a more engaging and interactive user experience.

\begin{figure}[tbp]
    \centering
    \includegraphics[width=0.45\textwidth]{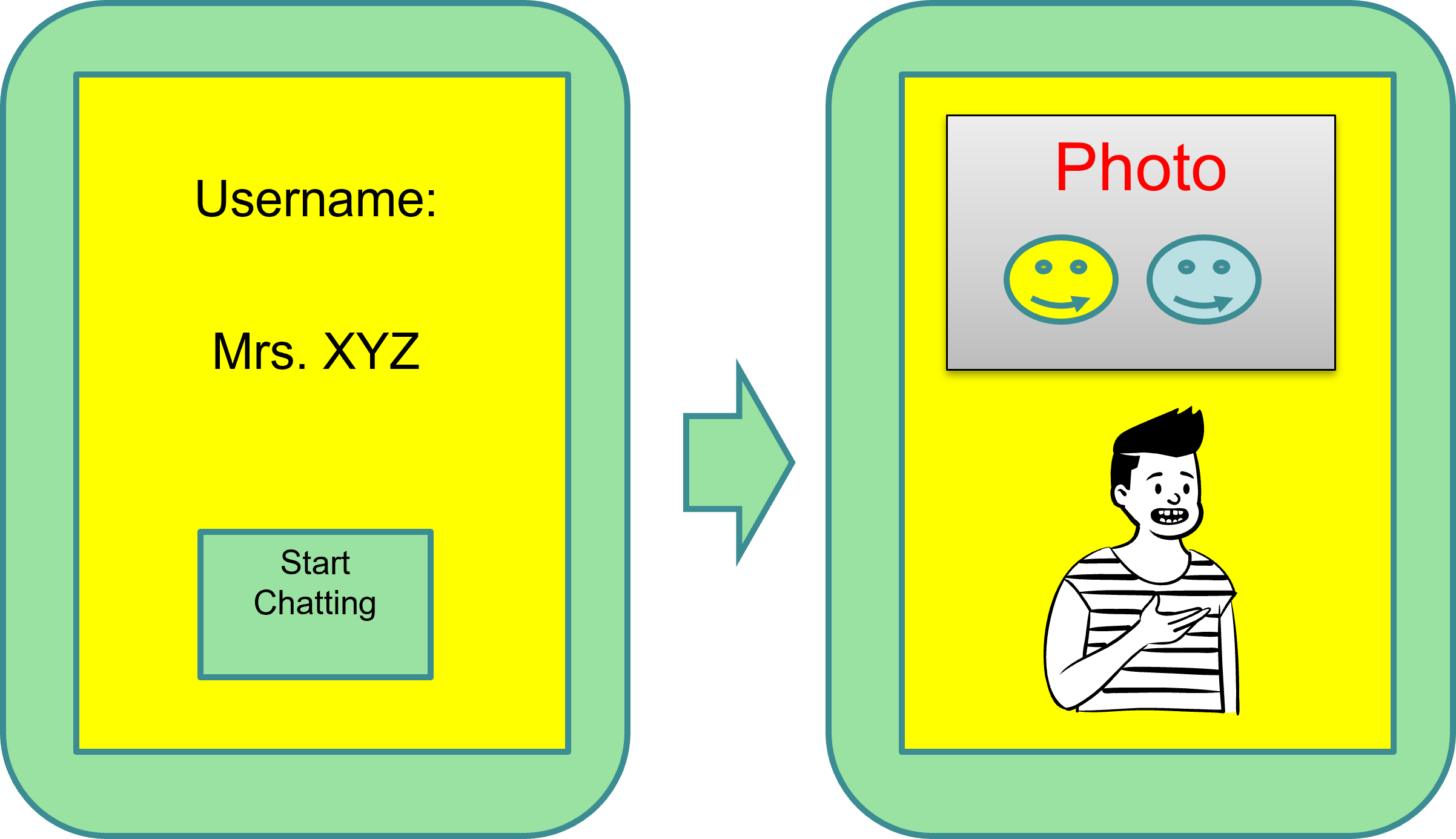}
    \caption{Mobile App Interface for start chatting and chat}
    \label{fig:ourtts}
\end{figure}

\begin{table*}[!htbp]
    \centering
    \begin{tabular}{|c|c|c|c||p{10cm}|}
        \hline
        Round & Role & Question & Agent Choice & Message \\ 
        \hline
        1 & Chatbot & Who &   & Do you recognize anyone in this photo? \\ 
        \hline
        2 & Elderly &   &  & Oh, this is my grandson! He’s always smiling, really adorable! \\ 
        \hline
        3 & Chatbot &   & C & Your grandson is so cute! Does this photo hold special memories? \\ 
        \hline
        4 & Elderly &   &  & Yes, this was taken during his first trip to Ocean Park! He loved watching the dolphin show. \\ 
        \hline
        5 & Chatbot &   & C & Ocean Park is so much fun! Does your grandson like any other animals? \\ 
        \hline
        6 & Elderly &   &  & He loves penguins! He always says they’re funny and even wants to bring one home! \\ 
        \hline
        7 & Chatbot & Who & D & Let’s go back to the photo—do you recognize anyone else in it? \\ 
        \hline
        8 & Elderly &  &   & Oh, besides my grandson, my daughter is also in the picture. They took it together! \\ 
        \hline
        9 & Chatbot & When & A & You answered correctly! When was this photo taken? \\ 
        \hline
        10 & Elderly &  & & It was taken last Christmas when we visited Ocean Park as a family. \\ 
        \hline
        11 & Chatbot &  & B & No worries, you might be mistaken. The previous question was about when the photo was taken, and the answer is Christmas Eve. Do you remember where it was taken? \\ 
        \hline
        12 & Elderly & Where &   & Oh, it was taken at home while we were preparing Christmas dinner. \\ 
        \hline
        13 & Chatbot &  What & A & You answered correctly! What were the people in the photo doing? \\ 
        \hline
        14 & Elderly &  &   & They were helping prepare Christmas dinner, chatting, and laughing—it was a joyful moment! \\ 
        \hline
        15 & Chatbot & Open-domain  & A & You answered correctly again! Anything else special about this photo? \\ 
        \hline
        16 & Elderly &   &   & I remember my grandson even sang a Christmas song that night—it was really beautiful! \\ 
        \hline
        17 & Chatbot &   & C & Christmas songs really bring festive vibes! Which song did your grandson sing? Does he enjoy singing other songs? \\ 
        \hline
        18 & Elderly &   &  & He sang *Jingle Bells* that night, and he also loves *Twinkle Twinkle Little Star*! \\ 
        \hline
        19 & Chatbot &   & D & Your grandson is truly talented—his singing makes Christmas even more warm and joyful! I hope he continues to enjoy music! Would you like to continue chatting about your grandson’s photos? \\ 
        \hline
        20 & Elderly &   &  & Sure! \\ 
        \hline
    \end{tabular}
    \caption{Sample interactions 1 between users and our chatbot}
    \label{tab:conversation}
\end{table*}

\begin{table*}[!htbp]
    \centering
    \begin{tabular}{|c|c|c|c||p{10cm}|}
        \hline
        Round & Role & Question & Agent Choice & Message \\ 
        \hline
        1 & Elderly &   &   & This dinner should have been enjoyable, but I really dislike my son-in-law now because we completely disagree on politics! \\ 
        \hline
        2 & Chatbot &   & C & Oh, political discussions can sometimes be challenging. How do you usually talk about these topics with your son? \\ 
        \hline
        3 & Elderly &   &   & We've tried discussing, but it always turns into an argument—he never listens to me! \\ 
        \hline
        4 & Chatbot &   & C & That sounds frustrating. Do you think there’s a way to make your conversations more constructive and peaceful? \\ 
        \hline
    \end{tabular}
    \caption{Sample interactions 2 between users and our chatbot}
    \label{tab:conversation}
\end{table*}

Caregivers can access this summary through the web portal, providing them with insights into the user’s preferred topics and conversational tendencies. This feature helps caregivers better understand engagement patterns, allowing for improved support and more meaningful interactions with the elderly user.

\subsection{Evaluation}

To assess the effectiveness of the proposed chatbot, we created a secondary LLM to simulate an elderly individual and engaged it in conversations with our chatbot. We designed the simulated elderly persona who likes to mention his grandson and with a tendency to frequently go off-topic, ensuring a realistic and dynamic interaction that mirrors natural conversational patterns.


Table V presents an example of a chatbot interacting with a simulated elderly user. The mobile app displays a family photo. The photo description of the photo is, "On Christmas Eve, the user had Christmas Eve dinner with his daughter at his daughter's home."  The chatbot begins by asking who is in the photo, as shown in conversation round 1.

If the user's response diverges from the expected answer, the chatbot engages in a natural conversation, as shown from rounds 3 to 6. In round 7, it gently redirects the discussion back to identifying the person in the photo. Upon receiving the correct answer, the chatbot responds positively and cheerfully before proceeding to the next question.

In round 11, when the user provides an incorrect response, the chatbot reiterates the previous question and offers the correct answer. This approach encourages the user to reflect on past memories and aids in memory recall. The chatbot then continues with the next question.

The final question in round 15 is open-ended, inviting the user to share additional thoughts or memories related to the event in the photo. After several rounds of conversation, in round 19, the chatbot is designed to conclude the discussion on the current photo.

To facilitate a smooth transition, the conversation history from rounds 1 to 18, along with relevant background parameters, is provided to the chat summary prompt in Table IV. The system updates the user’s profile, modifying: 
"Like = [calligraphy], Dislikes =[]" to "Like = [calligraphy, Ocean Park, grandchildren, penguins, Christmas]. Dislikes =[]". It also identifies the target person as the grandson. It also generates a summary of the chat as, "The elder shared his fond memories with his family, including his grandson’s first visit to Ocean Park to watch a dolphin show. He particularly liked penguins and even wanted to bring them home to raise. The photos recorded the time they were preparing dinner on Christmas Eve. Their grandson even sang Christmas carols, filling the whole family with a joyful atmosphere." After this process, the chatbot generates a response, asking whether the user would like to continue the conversation with a new photo featuring the target person, the grandson, ensuring a personalized and engaging interaction.

For another sample chat shown in Table VI, the summary is, "The user and his son-in-law have serious differences on political issues, which often turn discussions into arguments and affect the family atmosphere that should be relaxed. He hopes to find a more peaceful way of communication, but he still needs time to think calmly." The updated profile becomes:  {Like= [calligraphy], Dislike= [political disputes, son-in-law]}, demonstrating how the chat summary effectively refines user preferences in a structured and efficient manner.

This summary information is accessible through the portal, allowing caregivers to reference it as a meaningful topic for conversation during their next visit.

We used GPT-4 2024-12-01-preview as the backbone LLM. During the evaluation, we observed that the system’s reasoning in determining whether a user had directly answered a question was sometimes ambiguous. For instance, in round 3, the user responded to a question about their grandson, yet option C was selected because the response included additional details rather than a direct answer. Liu et al.\cite{b13} tested GPT-4’s logical reasoning ability across various benchmarks, finding that while humans typically achieve around 90\% accuracy, GPT-4's performance ranged between 60\% and 70\%. This discrepancy suggests that while GPT-4 excels at conversational engagement, its reasoning capabilities—especially in distinguishing response types—still present limitations. Future advancements in LLM models are expected to improve reasoning capabilities, addressing current limitations in distinguishing response types.









\section{Conclusion}

This study presents a personalized chatbot designed to facilitate meaningful conversations with elderly individuals. By employing goal-oriented dialogue, the chatbot engages users through discussions centered on family photos, fostering memory recall, cognitive stimulation, and positive reminiscence. The chatbot adapts to user preferences and identifies favored topics and individuals while dynamically adjusts interactions to sustain engagement. Its integration with a caregiver web portal further enhances practicality, allowing caregivers to upload photos, review conversations, and gain insights into users’ well-being.

This chatbot serves as a digital companion, reducing caregiver workload while emulating a social worker's approach to understanding individual preferences. Experimental results show that AI-driven dialogue flow improves conversational structure, ensuring a supportive and adaptive user experience.



\section*{Acknowledgment}

This work was supported by a grant from 
the Innovation and Technology Fund of the Hong Kong SAR, China (Project No: ITP/046/23LP).

\end{document}